\documentclass[journal]{IEEEtran}

\usepackage[utf8]{inputenc}
\usepackage[T1]{fontenc} 

\usepackage{amsmath,amsfonts,amssymb}
\usepackage{algorithmic}
\usepackage{algorithm}
\usepackage{array}
\usepackage[caption=false,font=normalsize,labelfont=sf,textfont=sf]{subfig}
\usepackage{textcomp}
\usepackage{stfloats}
\usepackage{url}
\usepackage{verbatim}
\usepackage{graphicx}

\usepackage{hyperref}
\usepackage{csquotes}
\usepackage{enumitem}
\usepackage{tikz}
\usetikzlibrary{shapes.geometric, arrows, arrows.meta, calc, fit}

\usepackage{colortbl}
\usepackage{xcolor}
\usepackage{multirow}
\usepackage{booktabs}
\usepackage{balance}

\usepackage[style=ieee, backend=biber,sorting=nty, mincitenames=1, maxcitenames=2]{biblatex}
\DefineBibliographyStrings{english}{%
  andothers = {et\addabbrvspace al\adddot},
}

\newcommand*{\enq}[1]{\enquote{{\itshape#1}}}

\definecolor{background-gray}{gray}{0.96}

\usepackage[framemethod=tikz]{mdframed}
\newmdenv[
  topline=false,
  bottomline=false,
  rightline=false,
  skipabove=\baselineskip,
  skipbelow=\baselineskip,
  innertopmargin=6pt,
  innerbottommargin=6pt,
  innerleftmargin=12pt,
  innerrightmargin=12pt,
  tikzsetting={draw=black, line width=4pt},
  linecolor=black,
  backgroundcolor=background-gray
]{results}

\newcommand{\fix}[1]{#1}

\begin{document}

\title{On the Need to Rethink Trust in AI Assistants for Software Development: A Critical Review}




\author{
\IEEEauthorblockN{Sebastian Baltes\IEEEauthorrefmark{2}\IEEEauthorrefmark{1}\thanks{\IEEEauthorrefmark{1}The first two authors contributed equally to this work.}, Timo Speith\IEEEauthorrefmark{2}\IEEEauthorrefmark{3}\IEEEauthorrefmark{1}, Brenda Chiteri\IEEEauthorrefmark{2}, Seyedmoein Mohsenimofidi\IEEEauthorrefmark{2},\\
Shalini Chakraborty\IEEEauthorrefmark{3}, Daniel Buschek\IEEEauthorrefmark{3}}\\
\IEEEauthorblockA{
\IEEEauthorrefmark{2}Heidelberg University, Germany\\
\IEEEauthorrefmark{3}University of Bayreuth, Germany\\
\IEEEauthorrefmark{2}\{sebastian.baltes, brenda.chiteri, s.mohsenimofidi\}@uni-heidelberg.de\\
\IEEEauthorrefmark{3}\{timo.speith, s.chakraborty, daniel.buschek\}@uni-bayreuth.de
}}


\maketitle

\begin{abstract}
Trust is a fundamental concept in human decision-making and collaboration that has long been studied in philosophy and psychology. 
However, software engineering (SE) articles often use the term \emph{trust} informally; providing an explicit definition or embedding results in established trust models is rare.
In SE research on AI assistants, this practice culminates in equating trust with the likelihood of accepting generated content, which, in isolation, does not capture the full conceptual complexity of trust.
Without a common definition, true secondary research on trust is impossible. 
The objectives of our research were: (1) to present the psychological and philosophical foundations of human trust, (2) to systematically study how trust is conceptualized in SE and the related disciplines human-computer interaction and information systems, and (3) to discuss limitations of equating trust with content acceptance, outlining how SE research can adopt existing trust models to overcome the widespread informal use of the term trust. 
We conducted a literature review across disciplines and a critical review of recent SE articles with a focus on trust conceptualizations. 
We found that trust is rarely defined or conceptualized in SE articles. Related disciplines commonly embed their methodology and results in established trust models, clearly distinguishing, for example, between \emph{initial trust} and \emph{trust formation} and between \emph{appropriate} and \emph{inappropriate trust}.
On a meta-scientific level, other disciplines even discuss whether and when trust can be applied to AI assistants at all.
Our study reveals a significant maturity gap of trust research in SE compared to other disciplines. We provide concrete recommendations on how SE researchers can adopt established trust models and instruments to study trust in AI assistants beyond the acceptance of generated software artifacts.
\end{abstract}


\section{Introduction}
\label{sec:introduction}

\begin{figure}
    \centering
    \includegraphics[width=1.0\linewidth]{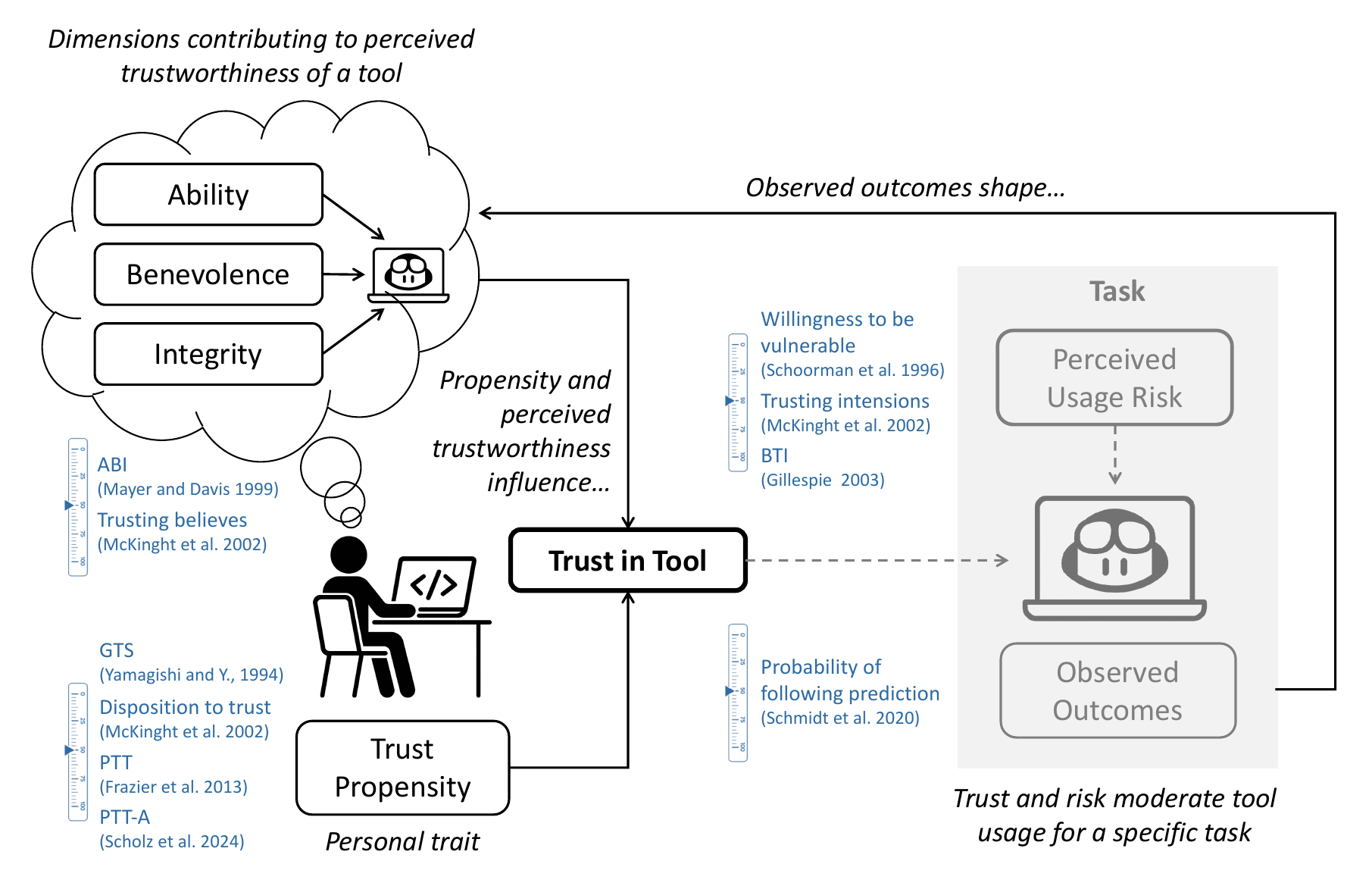}
    \caption{Trust model by \citeauthor{Mayer1995} applied to AI assistants with references to existing measurement instruments~\cite{MayerDavies1999, DBLP:journals/isr/McKnightCK02, Frazier01102013, Scholz17012025, DavidSchoorman02012016, gillespie201520, DBLP:journals/jds/SchmidtBT20}.}
    \label{fig:mayer-scales}
\end{figure}

Since the launch of ChatGPT in November 2022 and the subsequent hype around generative artificial intelligence (GenAI), AI assistants that help humans generate images, text, or source code have become an integral part of the daily work of many knowledge workers around the world \cite{mcgeorge2023chatgpt}.
GenAI tools, for example ChatGPT and GitHub Copilot, support software engineers in tasks such as understanding existing code, fixing bugs, or implementing new features~\cite{sridhara2023chatgpt,khojah2024beyond}.
GitHub claims that, as of February 2023, Copilot is \enq{behind an average of 46 percent of developers' code across all programming languages.}\footnote{\href{https://github.blog/ai-and-ml/github-copilot/github-copilot-now-has-a-better-ai-model-and-new-capabilities/}{GitHub Blog: GitHub Copilot now has a better AI model}}
The IDE integration of GitHub Copilot and similar AI development assistants allows tool vendors to collect fine-grained usage data to understand which suggestions developers accept, which they reject, and which they modify before committing a change~\cite{DBLP:journals/cacm/ZieglerKLRRSSA24}.

Previous work coauthored by employees of companies with commercial interests in increasing tool adoption, including Microsoft, Google, and Amazon, investigated what they refer to as \textbf{trust} in AI assistants~\cite{cheng2024would,DBLP:conf/fat/0004CF024,brown2024identifying, DBLP:journals/software/MurilloEDBKMM24, Sabouri2025, DBLP:conf/icse/JohnsonBFFZ23,}.
In some of those articles, the concept of \emph{trust} is operationalized as tool adoption in general or, more specifically, as a high \textbf{acceptance rate} for generated content.
In short, they equate trust with a high probability of developers accepting generated software artifacts, while ignoring \fix{aspects like} trust formation and the antecedents of trust (see \autoref{fig:mayer-scales}).
Compared to existing conceptualizations of trust in foundational disciplines such as psychology and philosophy, but also more applied disciplines such as human-computer interaction (HCI) and information systems (IS), we consider this view to be \textbf{too simplistic}.
It \textbf{disregards decades of research on trust}.

Although embedding software engineering (SE) research in existing models and theories of trust is not widespread \fix{(see \autoref{sec:literature-review})}, our community does have a history of discussing \textbf{trust in source code}.
For example, in his Turing Award lecture in 1984, Ken Thompson addressed \fix{trust in foreign code}, stating: \enq{You can't trust code that you did not totally create yourself... No amount of source-level verification or scrutiny will protect you from using untrusted code}~\cite{DBLP:journals/cacm/Thompson84}.
Given the widespread adoption of AI-based code generation and the persistent issue of hallucinations~\cite{DBLP:journals/corr/abs-2409-05746}, it becomes imperative to adopt a more \textbf{holistic perspective on trust}, one that extends beyond the notion of accepting AI-generated content.
Combining Ken Thompson's statement with the increasing amount of generated code, this leads to a critical question: \emph{Can developers genuinely trust generated software artifacts, or is there a conceptual difference between trust and mere acceptance of generated content?}

Our analysis of conceptualizations of trust in other disciplines reveals that viewing trust as mere acceptance of generated content is problematic for two key reasons. 
First, it \textbf{lacks a solid theoretical foundation} because it is unclear whether what is being measured is actually aligned with common conceptualizations of trust.
People's behavior, and hence their likelihood of accepting generated content, can be influenced not only by the perceived trustworthiness of the AI they are using, but also by other factors (see \autoref{fig:mayer-scales} for an example).
Factors that affect trust are not necessarily directly related to trust~\cite{Lee2004Trust, Mayer1995}.
Such factors include people's specific situation (e.g., time pressure), the type of AI system with which they are interacting, the specific task they are performing, and the potential consequences of decisions.

Second, viewing trust as mere acceptance of generated content is \textbf{potentially harmful}.
Even if trust could be measured through acceptance alone, this would not imply that the measured trust is appropriate trust.
Given the potential maintainability challenges posed by generated code~\cite{DBLP:journals/tosem/LiuLWTLLL24}, \fix{security risks associated with misplaced trust in AI-generated code~\cite{DBLP:conf/ccs/PerryS0B23, DBLP:conf/sp/PearceA0DK22, DBLP:conf/uss/SandovalPNKGD23, DBLP:conf/chi/SerafiniYN25, DBLP:conf/sac/KudriavtsevaHG25, DBLP:conf/wcre/MajdinasabBRDTK24, DBLP:conf/ccs/KlemmerHPLBPMRV24, DBLP:conf/sp/OhLPKK24}}, and the potential negative impact on developers' learning process~\cite{Yan2024, DBLP:journals/cacm/Shein24b}, researcher need to distinguish \textbf{appropriate} from \textbf{inappropriate trust}.
\fix{The fact that many developers distrust the accuracy of AI assistants\footnote{\href{https://survey.stackoverflow.co/2025/ai\#2-accuracy-of-ai-tools}{2025 Stack Overflow Developer Survey}} but nevertheless adopt these tools indicates that there is more to adoption than just trusting a tool.}

We advocate for a broader interdisciplinary approach to understanding trust in SE.
Our goal is to \textbf{motivate the SE research community to rethink how we conceptualize and measure trust}, especially in the context of AI assistants for software development.
\fix{We would like to establish that software cannot be developed with the goal of ``increasing trust'' without knowing what trust actually entails.
Without a scientifically validated trust model, trust becomes an empty marketing concept or, even worse, a concept used for \emph{ethics washing} (see \autoref{sec:trust-and-ai}).
Without an established trust model, there is no reference point to compare claims of ``increased trust'' to and check whether they are actually true.
More specifically, without a clear understanding of the concept of trust, it is difficult to consider trust in design goals, which in turn makes it difficult to design interactive AI assistants for software development in a way that addresses concrete trust-related challenges.}

Note that in this paper, we focus on \textbf{human or human-like trust}, not trust in a more technical sense (e.g., \fix{the implementation of} chains of trust in a security context).
Our target audience are SE researchers planning to design studies on trust as well as editors and reviewers of SE research who are responsible for ensuring our communities' epistemological and methodological standards when it comes to researching interdisciplinary topics such as trust.
To start a community discourse on rethinking trust and to support researchers who want to adopt and adapt established trust models, we address the following research questions (RQs):

\begin{description}[style=multiline, leftmargin=10mm]
\item[RQ1] How is trust conceptualized in foundational disciplines and applied disciplines related to SE?
\item[RQ2] How is trust conceptualized in SE?
\item[RQ3] How can we adopt existing trust models for SE research to study trust in AI assistants? 
\end{description}

After introducing the foundations of human trust based on research in psychology and philosophy (\autoref{sec:foundationoftrust}), we present the results of a literature review (\autoref{sec:literature-review}) on conceptualizations of trust in HCI and IS (\textbf{RQ1}) as well as SE (\textbf{RQ2}).
We complement that literature review with a critical review of trust conceptualizations in recent SE articles (\textbf{RQ2}) and then synthesize the results of the two literature reviews into a discussion on how to rethink trust in AI assistants for software development (\textbf{RQ3}), including actionable recommendations for SE researchers (\autoref{sec:se-adoption}).
We then discuss threats to validity (\autoref{sec:validity}) and conclude the paper (\autoref{sec:conclusion}).

\section{Background: Foundations of Trust}
\label{sec:foundationoftrust}
\emph{Trust} and the related concept of \emph{trustworthiness} have been discussed and conceptualized long before disciplines such as SE, HCI, or IS recognized these concepts. 
Therefore, before we dive into our cross-disciplinary literature review of trust, we outline how philosophy and psychology, that is, disciplines that have been studying the foundations of human behavior for a long time, conceptualize \emph{human trust}.

First, it is important to distinguish \emph{trust} from \emph{trustworthiness}.
Both philosophy and psychology agree that \emph{trust} is an attitude that one entity (the trustor) has toward another entity (the trustee) that it hopes will be \emph{trustworthy}.
Hence, leaving \emph{perceived trustworthiness} aside, \emph{trustworthiness} is considered a property of the trustee and not an attitude \cite{McLeod2023Trust, Lee2004Trust, 10.1145/3696449}. 
Furthermore, both disciplines agree that \emph{trust} \fix{and \emph{trustworthiness} are contextual matters}: a trustor may trust a trustee in one situation or for one specific task while, at the same time, not trusting them in another situation or for another task.
Typically, the trustor is human, while it is debatable whether the trustee must also be human (see \autoref{sec:philosophy}).

\subsection{Psychology}
\label{sec:psychology}

\tikzstyle{box} = [rectangle, rounded corners, minimum width=2.5cm, minimum height=1cm, text width=2.5cm, text centered, draw=black, fill=white]
\tikzstyle{arrow} = [thick,->,>=stealth]

\begin{figure}[tb]
    \centering
    \scalebox{0.55}{
\begin{tikzpicture}[node distance=1.75cm]

\node (ability) [box] {Ability};
\node (benevolence) [box, below of=ability] {Benevolence};
\node (integrity) [box, below of=benevolence] {Integrity};

\node (trustworthiness) [draw, dashed, fit={(ability) (integrity)}, inner sep=0.5cm, label=above:{Factors of Perceived Trustworthiness}] {};

\node (trust) [box, right of=benevolence, xshift=2cm] {Trust};
\node (propensity) [box, below of=trust, xshift=-1.25cm, yshift=-1.75cm] {Trustor's Propensity};
\node (perceivedrisk) [box, above of=trust, xshift=1.75cm] {Perceived Risk};
\node (risktaking) [box, right of=trust, xshift=2cm] {Risk Taking in Relationship};
\node (outcomes) [box, right of=risktaking, xshift=2cm] {Outcomes};

\draw [arrow] (ability.east) -- coordinate[pos=0.6] (at) (trust.north west);
\draw [arrow] (benevolence.east) -- coordinate[pos=0.6] (ab) (trust.west);
\draw [arrow] (integrity.east) -- coordinate[pos=0.6] (ai) (trust.south west);

\draw [arrow] (propensity.north) -- (trust.south);
\draw [arrow] (propensity.north) -- (at.south);
\draw [arrow] (propensity.north) -- (ab.south);
\draw [arrow] (propensity.north) -- (ai.south);

\draw [arrow] (trust.east) -- coordinate[pos=0.4] (pr) (risktaking.west);
\draw [arrow] (perceivedrisk.south) -- (pr.north);
\draw [arrow] (risktaking.east) -- (outcomes.west);

\draw [arrow] (outcomes.south) -- ++(0,-4cm) -- ++(-11.25cm,0) -- (trustworthiness.south);
\end{tikzpicture}}
\caption{Trust model by \citeauthor{Mayer1995} (own illustration based on~\cite{Mayer1995}).}
\label{fig:abi_model}
\end{figure}

In psychology, two models have been very influential in the discourse on trust.
Those models have also been adopted in other fields.
In \citeyear{Mayer1995}, \citeauthor{Mayer1995}, surveying the literature on trust, created the \emph{ability, benevolence, and integrity model of trust}~\cite{Mayer1995} (see \autoref{fig:abi_model}). 
They defined trust as \emph{\enq{the willingness of a party to be vulnerable to the actions of another party based on the expectation that the other will perform a particular action important to the trustor, irrespective of the ability to monitor or control that other party}} \cite[p.~712]{Mayer1995}.

In this model, \emph{ability} refers to the trustee's competence in a given domain, \emph{benevolence} captures the trustee's (assumed) goodwill and positive intentions toward the trustor, and \emph{integrity} refers to the trustee's adherence to ethical principles the trustor agrees with. The model suggests that trust is high when all these three components of trustworthiness are perceived by the trustor as high, while deficiencies in any factor can weaken trust. How much (changes in) the factors affect trust depends on the trustor's \emph{propensity to trust}. This propensity also influences the trustor's \emph{initial level of trust}. Trust can manifest itself in specific situations, depending on the level of trust and the \emph{risk} perceived by a trustor when they trust the trustee to perform a certain task.
Depending on the outcome of this risk-taking, the perception of the three factors can change. 

\begin{figure}[tb]
    \centering
    \large
    \resizebox{\columnwidth}{!}{%
    \begin{tikzpicture}[
  x=1.0cm,y=1.0cm,
  >=Stealth,
  axis/.style={line width=3.2pt},
  dashshape/.style={draw=black,line width=1.25pt,dash pattern=on 4.2pt off 3.0pt,fill=gray!35},
  dashline/.style={draw=black,line width=1.25pt,dash pattern=on 4.2pt off 3.0pt},
  dotline/.style={draw=black,line width=1.0pt,densely dotted},
  callout/.style={draw=black,line width=1.3pt},
  txt/.style={align=left,text width=6.0cm},
]

\def\W{10.0}      
\def\H{9.75}      

\def\Lytop{3.58}
\def\Lybot{3.14}
\def\Lxout{3.72}
\def\Lxin{3.24}
\path[dashshape]
  (0,\Lybot) -- (\Lxin,\Lybot) -- (\Lxin,0) -- (\Lxout,0) -- (\Lxout,\Lytop) -- (0,\Lytop) -- cycle;

\def\Bytop{6.75}
\def\Bybot{6.33}
\def\Bxend{6.93}
\path[dashshape] (0,\Bybot) rectangle (\Bxend,\Bytop);

\def\Wy{6.27}
\def\WxtopL{6.33}
\def\WxtopR{6.93}
\def\WxbaseL{5.57}
\def\WxbaseR{7.76}
\draw[dashline] (\Bxend,\Bybot) -- (\WxtopR,\Wy);
\path[fill=gray!35] (\WxtopL,\Wy) -- (\WxtopR,\Wy) -- (\WxbaseR,0) -- (\WxbaseL,0) -- cycle;
\draw[dashline] (\WxtopL,\Wy) -- (\WxbaseL,0);
\draw[dashline] (\WxtopR,\Wy) -- (\WxbaseR,0);
\draw[dashline] (\WxbaseL,0) -- (\WxbaseR,0);

\pgfmathsetmacro{\m}{\Lytop/\Lxout}
\coordinate (O) at (0,0);
\draw[dotline] (O) -- ({0.92*\W},{0.92*\W*\m});

\draw[axis] (0,0) -- (0,\H);
\draw[axis] (0,0) -- (\W,0);

\node[font=\bfseries\Large] at (-0.95,0.5*\H) {Trust};
\node[font=\bfseries\Large,align=center] at (0.5*\W,-0.95)
  {Automation Capability\\(Trustworthiness)};

\node[txt,anchor=west] at (2.25, 8.5)
  {\textbf{Overtrust:} Trust exceeds\\system capabilities,\\leading to misuse};

\node[txt,anchor=west] (caltxt) at (9.25,7)
  {\textbf{Calibrated trust:} Trust\\matches system capabilities,\\leading to appropriate use};

\node[txt,anchor=west] at (8, 4)
  {\textbf{Distrust:} Trust falls short\\of system capabilities,\\leading to disuse};

\node[txt,anchor=east] at (1,1.25)
  {\textbf{Good resolution:} A range\\ of system capability maps \\ onto the same range of trust};

\node[txt,anchor=west] at (9.25,1.25)
  {\textbf{Poor resolution:} A large range\\of system capability maps onto\\a small range of trust};

\coordinate (goodMarkTop) at (0.5, 2.25);
\coordinate (goodMarkBot) at (0.5, 0.25);
\draw[callout] (goodMarkBot) -- (goodMarkTop);
\draw[-{Stealth[length=7pt,width=8pt]},callout]
  ($(goodMarkTop)+(0.0,-0.5)$) -- (3.55,0.5);

\coordinate (poorMarkTop) at (9.25, 2.25);
\coordinate (poorMarkBot) at (9.25, 0.25);
\draw[callout] (poorMarkBot) -- (poorMarkTop);
\draw[-{Stealth[length=7pt,width=8pt]},callout]
  ($(poorMarkTop)+(0.0,-0.5)$) -- (6.5,0.5);
  
\coordinate (calBracketTop) at (9.25, 8);
\coordinate (calBracketBot) at (9.25, 6);
\draw[callout] (calBracketBot) -- (calBracketTop);
\draw[-{Stealth[length=7pt,width=8pt]},callout]
  ($(calBracketTop)+(0.0,-0.5)$) -- (8.4,8);
  
\end{tikzpicture}
    }
    \caption{Trust model by \citeauthor{Lee2004Trust} \fix{(own illustration based on~\cite{Lee2004Trust})}.}
    \label{fig:trust_model_leeandsee}
\end{figure}

\citeauthor{Lee2004Trust} take this model, expand it and apply it to automation contexts~\cite{Lee2004Trust}.
They introduce the concepts of \emph{calibration}, \emph{resolution}, and \emph{specificity} (see \autoref{fig:trust_model_leeandsee}).
\emph{Calibration} concerns the crucial question of whether the trust a trustor has corresponds to the real capabilities, that is, the actual trustworthiness of a system.
Ill-calibrated trust may lead to disuse (in the case of \emph{dis-/undertrust}) or misuse (in the case of \emph{overtrust}) of a system.
\textit{Resolution} is about how changes in system capabilities change levels of trust.
For example, with low resolution, high changes in capability would lead to small changes in trust.
Finally, \emph{specificity} reflects how granular the trust is.
Specificity emphasizes the context-dependence of trust by stating how much trust depends on the capabilities of a particular system or the situation.
According to \citeauthor{Lee2004Trust}, good calibration, high resolution, and high specificity of trust can mitigate misuse and disuse of automation systems and enhance human-automation partnerships. 
Although they also introduce a model on the interaction of context, agent characteristics, and cognitive properties with the appropriateness of trust, their work is mostly cited for making the above distinctions.

\subsection{Philosophy}
\label{sec:philosophy}

\citeauthor{McLeod2023Trust} has analyzed philosophical debates on trust and distills three requirements for trust~\cite{McLeod2023Trust}. In summary, trust requires that one can:
\begin{enumerate}
    \item be \emph{vulnerable} to others, in particular to betrayal,
    \item rely on others to be \emph{competent} to do what one wishes to trust them to do, and
    \item rely on them to be \emph{willing} to do it.
\end{enumerate}

For most philosophers, trust is a kind of \emph{reliance}, although it involves \enq{some extra factor} \cite{McLeod2023Trust}. 
This extra factor is often determined by how philosophers interpret the three requirements listed above.

For example, regarding the factor of \emph{vulnerability to betrayal}, trust is often \fix{connected to} \textit{anti-monitoring} (see, e.g., \cite{Baier1986Trust, Keren2014Trust}).
The idea here is that if a person monitors an entity, this person cannot be betrayed by that entity.
At most, the person could be disappointed if the entity does not perform as intended.
Here, there is an interesting difference to \citeauthor{Mayer1995}'s theory of trust, which explicitly decouples monitoring from trust.
However, their theory is flexible enough to \fix{capture the same intuition concerning} monitoring situations.
For example, an employee may devalue the \emph{benevolence} of (and thus the trust in) their employer due to constant monitoring.

The factors of \emph{competence} and \emph{willingness} (i.e., motivation) have received the most attention in the debate, with many competing theories that exist~\cite{McLeod2023Trust}.
These theories range from the trustee acting out of self-interest~\cite{Hardin2002Trust} or goodwill~\cite{Jones1999Second-Hand, Baier1986Trust} to theories of \fix{networks of trust}~\cite{Frost2014Imposters}.
In summary, while there is a minimum consensus of the three requirements mentioned above, philosophers differ so much in the exact formulations of these requirements that some even assume that there are different types of trust.
This would mean that trust is not one form of reliance but \fix{several} different ones~\cite{Simpson2012Trust, Scheman2020Trust, chen2021trust}. 

Philosophers place more emphasis on the \emph{preconditions of trust} and the \fix{conditions} of its \emph{formation} than psychologists.
Therefore, in addition to the concept of \emph{calibrated trust} (which they call \emph{well-grounded} trust), they also introduce the concepts of \emph{justified} and \emph{plausible} trust \cite{McLeod2023Trust}.
\emph{Justification} is about the indicators that have been consulted for trust formation, that is, whether it is based on good evidence.
If a person happens to trust the system in the right way, but does so by chance without any indications, then philosophers would not speak of justified trust. 
Finally, \emph{plausibility} is about whether the preconditions for trust are satisfied.
For example, if two parties are antagonistic, there can be no trust between them.

\emph{Justified} trust and \emph{well-grounded} trust are independent of each other. 
A person might be justified in their trust, for instance, by receiving reports on the trustee's performance.
However, these reports might be forged, significantly overestimating actual performance, so that the ensuing trust would be ungrounded.
Similarly, the above example of the trustor trusting the trustee to the right level by chance is a case of well-grounded but unjustified trust.

\subsection{Philosophy and Psychology on Trust and AI}
\label{sec:trust-and-ai}

Philosophical and psychological discussions about trust continue to this day, often in specific areas of application.
AI is one of them, especially in recent times.
There are \fix{three} discussions that are worth highlighting here.
In particular, with regard to the \emph{plausibility of trust}, there are discussions as to whether it makes sense to apply the term trust to AI at all.
\fix{From a philosophical perspective, current AI is not conscious, so it does not make sense to speak of it `betraying' us.}

This discussion even made it into the media when the \emph{Guidelines for Trustworthy AI} of the high-level AI expert group set up by the European Commission were published. 
One of the members of this group, the German philosopher Thomas Metzinger, gave an interview shortly before the guidelines were published, in which he complained about the strong involvement of companies in the design of the guidelines. He claimed that these would use the principle of \emph{trustworthy AI} for \emph{ethics washing}, because in his opinion, the term trustworthy AI is conceptual nonsense.\footnote{\href{https://www.tagesspiegel.de/politik/ethics-washing-made-in-europe-5937028.html}{EU guidelines: Ethics washing made in Europe (Tagesspiegel)}} 
While some authors agree with Metzinger in that the concept of trust cannot be applied to AI systems, for example, since they do not possess emotive states or cannot be held responsible~\cite{Ryan2020AI, Dorsch2024Quasi}, others defend the possibility to trust AI, for instance, by showing how philosophical theories of trust can be adapted to AI~\cite{Viehoff2023Making, Safdari2024Toward, DBLP:conf/fat/JacoviMMG21, chen2021trust, DBLP:journals/ethicsit/Nickel22} or by differentiating trust in AI from trust in humans~\cite{Zanotti2023Keep} (see \cite{duran2025trust} for an overview of the debate).

\fix{This leads to a second, more practice-oriented discussion: How is trustworthiness operationalized?}
Irrespective of the discussion as to whether the concept of trust can be meaningfully applied to AI, the \emph{Guidelines for Trustworthy AI} had strong effects on the subsequent \fix{operationalization of AI trustworthiness (and with that trust in AI)}. By specifying requirements that are essential for trustworthiness\fix{, e.g. technical robustness, safety, transparency, and accountability,} 
they provide fertile ground for psychological and philosophical research to operationalize these requirements more precisely (see, e.g., \cite{mattioli2024overview, li2024developing}).
\fix{Encompassing some of these requirements, especially technical robustness, \emph{computational reliability} came to be an important factor for trustworthiness in philosophical discussions (see, e.g., \cite{Ferrario2022Explainability}).}

The \fix{last} noteworthy discussion revolves around the relationship between explainable AI (XAI) and trust in AI.
Increasing trust is one of---if not the---primary goal of XAI~\cite{Kaestner2021Relation, Langer2021What}.
\citeauthor{Kaestner2021Relation} refer to this as the \emph{explainability-trust hypothesis}~\cite{Kaestner2021Relation}.
However, studies show that providing explanations does not consistently increase trust, and in many cases, even decreases it \cite{Kaestner2021Relation, Langer2021What}.
Here, trust calibration plays an important role, as explanations that reveal system errors should lead to lower trust.
Unfortunately, psychological biases influence this process, as people sometimes trust a system simply because of the illusion of an explanation~\cite{DBLP:conf/chi/EibandBKH19}.
On the same topic, \citeauthor{Ferrario2022Explainability} argue, drawing on the concept of anti-monitoring, that XAI cannot directly contribute to trust, as it constitutes a form of monitoring~\cite{Ferrario2022Explainability}. \fix{However, they argue that XAI can indirectly support trust because human-AI interaction that is supported by XAI is generally more trustworthy.}

Regarding biases and the possibility of generating problematic content, there is now more and more research that argues that, for a system to be adequately trusted, it is not just the system that needs to function correctly.
Rather, for a system to be perceived as benevolent, there must be regulation, auditing processes, and established standards that govern the entire engineering process \cite{winfield2018ethical, manzini2024should, doi:10.1177/00218863241285033}. \citeauthor{winfield2018ethical}, for instance, argue that ethical governance, consisting of ethical codes of conduct, ethics training, responsible innovation, and transparency, is essential to build trust in AI systems~\cite{winfield2018ethical}.
Similarly, \citeauthor{manzini2024should} argue that user trust requires evidence on the functionality of an AI assistant at the level of its design, the practices of the developing organization, and oversight by external bodies~\cite{manzini2024should}.

\begin{results}
\fix{\textbf{RQ1} \textsc{(trust conceptualizations)}:\\
In \textbf{psychology}, the \emph{ability, benevolence, and integrity} model of trust is common.
How much (changes in) these factors affect trust depends on the trustor's \emph{propensity to trust}, which again affects the trustor's \emph{initial level of trust}.
A central aspect is the \emph{risk} perceived by a trustor when they trust the trustee to perform a certain task.
The model was later extended to cover \emph{calibration}, \emph{resolution}, and \emph{specificity} of trust.
E.g., ill-calibrated trust may lead to \emph{disuse} (in the case of \emph{undertrust}) or \emph{misuse} (in the case of \emph{overtrust}) of a system.\\
In \textbf{philosophy}, \emph{anti-monitoring} (not being able to monitor the trustee) is essential.
Philosophers place more emphasis on the \emph{preconditions of trust} and the conditions of \emph{trust formation} than psychologists.
They refer to \emph{calibrated trust} as \emph{well-grounded trust} and add the concepts \emph{justification} and \emph{plausibility} of trust. 
There are ongoing discussions whether trust can be meaningfully applied to AI.}
\end{results}

\section{Literature Review (RQ1 and RQ2)}
\label{sec:literature-review}

After outlining the foundations of trust, we now turn to the methodology and results of our review of trust conceptualizations across the disciplines IS, HCI, and SE.

\subsection{Methodology}
\label{sec:method}

The goal of our research was to understand how trust is conceptualized in SE (\textbf{RQ2}) compared to other disciplines (\textbf{RQ1}) and, based on these conceptualizations, to explore how existing trust models can be adapted to study AI assistants in software development (\textbf{RQ3}).
To achieve this, we conducted a literature review across disciplines.

\subsubsection{\textbf{Literature review across disciplines}}

We decided to use \emph{Google Scholar} for the literature review, because portals such as the \emph{ACM Digital Library} and the \emph{IEEE Xplore Digital Library}, which are commonly used for literature reviews in SE, do not cover common publication venues in IS.
We used the Python library \emph{scholarly} to automate the retrieval of articles using a common search query across disciplines: 

{\small
\begin{verbatim}
FIELD (trust OR trustworthiness OR trustworthy)
\end{verbatim}
}

The values for FIELD were `software engineering', `human-computer interaction', and `(management OR information systems)'.
We selected HCI and IS because they are closely related to SE, but we expected them to have a more mature view on trust and its conceptualizations, in particular because both HCI and IS have a strong connection to psychology.
For the IS search queries, we added `management'  because information systems and management have overlapping publication venues.
In fact, the most reputable IS venue is called \emph{Management Information Systems Quarterly} (see the rankings linked from the inclusion criteria listed below).
Since our goal was to apply existing trust models to AI assistants, we further retrieved AI-specific articles by adding the following to the before-mentioned queries:

{\small
\begin{verbatim}
(ai OR genai OR artificial intelligence)
\end{verbatim}
}

For each discipline, we created two lists based on these queries (e.g., \textsc{HCI} and \textsc{HCI-AI}, see \autoref{tab:articles-literature-review})
and retrieved 50 articles from each list, selecting those with the highest search rank as of March 2025, resulting in a total of 300 articles to review.
We then identified and marked duplicates across the lists.
We assigned one author to each discipline who then reviewed the articles according to the inclusion and exclusion criteria we defined.
We organized weekly meetings to discuss unclear cases and applied the following criteria for the literature review across disciplines that we conducted to address \textbf{RQ1} and \textbf{RQ2}: 

\begin{itemize}[leftmargin=*]
\item \textbf{Include} papers published in the target discipline (use common rankings such as \emph{ICORE}\footnote{\url{https://portal.core.edu.au/conf-ranks/}} for SE or the \emph{ABDC JQL}\footnote{\url{https://abdc.edu.au/abdc-journal-quality-list/}} for IS; do not filter by rating, only exclude venues that are not listed in discipline-specific rankings). 
\item \textbf{Include} papers published in the main or research tracks of conferences and regular journal papers. Include literature reviews and special issues on human trust (analyzed separately, see \autoref{sec:literature-reviews-special-issue}).
\item \textbf{Exclude} other editorials, short papers ($<5$ pages), vision papers, textbooks, and dissertations.
\item \textbf{Exclude} papers that do not discuss human/human-like trust but trust in a technical sense (e.g., in a security context). 
\end{itemize}

For SE, we included 13 unique primary research papers (one paper was present in both lists), seven literature reviews (of which two were also part of the IS list; only three were SE-specific), and one special issue that overlaps with IS.
For HCI, we included 22 unique primary research papers (five papers were present in both lists) and five literature reviews (two present in both lists; two overlapping with IS).
For IS, we included 24 unique primary research papers (no overlap), four literature reviews (two also present in the SE lists, three without a strong IS focus), and three special issues on trust (one overlapping with SE).
Among the excluded SE papers were two HCI papers, which we added to the list of HCI papers (hence we have 24 HCI research papers in total).
Similarly, the HCI list contained one IS paper, which we added to the list of IS papers (hence we have 25 IS research papers in total).
Across the SE, HCI, and IS lists, we identified 15 articles that were published in philosophy venues (including ethics and fairness) or psychology venues. We read those papers and either integrated them into the background (\autoref{sec:foundationoftrust}) or the discussion (\autoref{sec:se-adoption}) section.

\autoref{tab:articles-literature-review} lists all the included papers.
Row \textsc{PH+PS} lists all philosophy and psychology papers that we integrated as mentioned above.
For the other papers, we downloaded the corresponding PDF files, searched for all occurrences of the substring ``trust,'' and extracted all paragraphs that conceptualized trust either directly or indirectly.
We also kept track of which trust models the papers referred to (if any) and whether they customized (adapted or extended) existing trust models.
We report the results of this analysis in \autoref{sec:trustacrossdisciplines}.

Many of the excluded SE papers were either not full papers or published in non-SE venues.
This, in conjunction with the fact that only few SE papers contained explicit conceptualizations of trust, contributed to the decision to follow up with a critical review of recently published articles on trust in leading SE venues (see below).

\subsubsection{\textbf{Critical review of recent SE literature}}

A critical review focuses on evaluating key issues and analyzing a sample of studies with shared characteristics~\cite{baltes2022sampling}.
The issue we intended to evaluate is the lack of trust conceptualization in SE research (related to \textbf{RQ2}).
To this end, we used \emph{dblp SPARQL} to retrieve all articles that met the inclusion/exclusion criteria: 

\begin{itemize}[leftmargin=*]
\item \textbf{Include} papers in leading SE conferences (ICSE, ASE, FSE) and journals (TSE, TOSEM) published recently, i.e., just before or after the GenAI hype (2022--2024).
\item \textbf{Include} papers published in the main or research tracks of conferences and regular full journal papers.
\item \textbf{Exclude} editorials, special issues, short papers, vision papers, papers from co-located events, etc.
\end{itemize}

For the 2670 papers that met our inclusion criteria, we retrieved all abstracts from the ACM and IEEE digital libraries and searched for the substring ``trust'' in the title and abstract.
Of the 2670 papers, 58 mentioned trust.
We downloaded the corresponding PDF files and manually searched for and extracted implicit and explicit conceptualizations of trust.
We report the results of this analysis in \autoref{sec:criticalreviewinse}.

\subsubsection{\textbf{Data availability}}

We provide all retrieval and analysis scripts and all metadata we retrieved from \emph{Google Scholar} and \emph{dblp} as part of our supplementary material~\cite{baltes_2025_15130575}.
Moreover, we share the annotated lists of papers, including exclusion reasons, and all extracted excerpts.
All included research papers are referenced in \autoref{tab:articles-literature-review}.

\subsection{Trust Conceptualizations across Disciplines}
\label{sec:trustacrossdisciplines}

In this section, we report the results of our literature review across disciplines.
Note that with this review, our goal was not to achieve a complete coverage of all articles that describe conceptualizations of trust in the target disciplines.
Our goal was to identify whether articles in the selected disciplines commonly refer to established trust models, with the goal of comparing SE articles to related disciplines.
Although we acknowledge this as a limitation of our research, the difference between the SE articles and the HCI/IS articles was so clear that our observations would most likely not change considerably when scaling up the analysis. 
\autoref{tab:trust-models} lists the trust models that at least two articles referred to.
The models proposed by \citeauthor{Mayer1995}~\cite{Mayer1995} and \citeauthor{Lee2004Trust}~\cite{Lee2004Trust} are used across HCI and IS, while \citeauthor{DBLP:journals/jsis/McKnightCK02}~\cite{DBLP:journals/jsis/McKnightCK02} and \citeauthor{DBLP:journals/misq/GefenKS03}~\cite{DBLP:journals/misq/GefenKS03} are IS-specific.
The other models listed in the table are primarily referenced in HCI, but less frequently than the \citeauthor{Mayer1995}~\cite{Mayer1995} and \citeauthor{Lee2004Trust}~\cite{Lee2004Trust} models.

\begin{table*}
  \caption{Common trust models across disciplines.}
  \label{tab:trust-models}
  \vspace{-0.25\baselineskip}
  \centering
  \begin{tabular}{|p{0.1\linewidth}|p{0.22\linewidth}|p{0.14\linewidth}|p{0.12\linewidth}|p{0.12\linewidth}|p{0.11\linewidth}|}
    \hline \rowcolor{gray!40}
    \textbf{Trust Model} & \textbf{Summary} & \textbf{HCI} & \textbf{HCI-AI} & \textbf{IS} & \textbf{IS-AI} \\
    \hline
    
    Mayer~et~al.\newline(1995)~\cite{Mayer1995} & Multi-disciplinary model of organizational trust. & \cite{DBLP:journals/computer/SousaLCM24, DBLP:journals/behaviourIT/Gulati0L19, DBLP:journals/ijmms/RiegelsbergerSM05, pinto2022trust, DBLP:journals/ijmms/CorritoreKW03a, Cai02012025, DBLP:conf/compsac/NothdurftBLM12} & \cite{DBLP:journals/computer/SousaLCM24, DBLP:journals/behaviourIT/Gulati0L19, DBLP:journals/pacmhci/BanovicYRL23, pinto2022trust} & \cite{DBLP:journals/jsis/LiHV08, DBLP:journals/ejis/SollnerHL16, Zhang2018, Tansley2007} & \cite{DBLP:journals/electronicmarkets/ThiebesLS21, DBLP:journals/jds/SchmidtBT20, Dennis03042023, info:doi/10.2196/15154} \\
    \hline \rowcolor{gray!10}
    
    Lee\&See\newline(2004)~\cite{Lee2004Trust} & Trust in automation systems and technology. & \cite{DBLP:journals/computer/SousaLCM24, pinto2022trust, Cai02012025, DBLP:conf/compsac/NothdurftBLM12, DBLP:conf/chi/WidderDHHD21} & \cite{DBLP:journals/computer/SousaLCM24, DBLP:journals/pacmhci/BanovicYRL23, DBLP:journals/chb/SuenH23, pinto2022trust, DBLP:journals/ijhci/YokoiEFN21} & \cite{DBLP:journals/ejis/SollnerHL16} & \cite{DBLP:journals/electronicmarkets/ThiebesLS21, info:doi/10.2196/15154} \\ 
    \hline
    McKnight~et~al.\newline(2011)~\cite{DBLP:journals/tmis/McKnightCTC11} & Model of trust in technology grounded in system-level traits.
    & \cite{DBLP:journals/computer/SousaLCM24, DBLP:journals/behaviourIT/Gulati0L19} & \cite{DBLP:journals/computer/SousaLCM24, DBLP:journals/behaviourIT/Gulati0L19} & None & \cite{DBLP:journals/electronicmarkets/ThiebesLS21} \\ 
    \hline \rowcolor{gray!10}

    Gulati~et~al.\newline(2019)~\cite{DBLP:journals/behaviourIT/Gulati0L19} & Empirically validated model for measuring intelligent systems trust. & \cite{DBLP:journals/computer/SousaLCM24, pinto2022trust} & \cite{DBLP:journals/computer/SousaLCM24, pinto2022trust} & None& None \\ 
    \hline

    McAllister\newline(1995)~\cite{mcallister1995affect} & Dual-factor model of organizational interpersonal trust.  & \cite{DBLP:conf/compsac/NothdurftBLM12, DBLP:journals/ijmms/RiegelsbergerSM05} & \cite{DBLP:journals/pacmhci/BanovicYRL23} & None & None \\ 
    \hline \rowcolor{gray!10}
    
    Rousseau~et~al.\newline(1998)~\cite{rousseau1998not} & Integration of organizational trust across disciplines. & \cite{DBLP:journals/ijmms/RiegelsbergerSM05, Cai02012025} & None & \cite{DBLP:conf/hicss/ChopraW03} & None \\ 
    \hline
    
    Madsen\&Gregor\newline(2000)~\cite{madsen2000measuring} & Human-computer trust model, focus on affective \& cognitive trust.& \cite{DBLP:conf/compsac/NothdurftBLM12, DBLP:conf/um/KrausWM20} & \cite{DBLP:journals/pacmhci/BanovicYRL23} & None& None \\ 
    \hline \rowcolor{gray!10}
    McKnight~et~al.\newline(2002)~\cite{DBLP:journals/jsis/McKnightCK02} & Model of initial trust in e-commerce settings. & None & None & \cite{DBLP:journals/jsis/LiHV08, DBLP:journals/tem/ThatcherMBAR11, DBLP:conf/hicss/ChopraW03, DBLP:journals/jcis/HoehleHG12, DBLP:journals/db/LiHV06} & \cite{DBLP:journals/isci/VirvouTT24} \\
    \hline 
    
    Gefen~et~al.\newline(2003)~\cite{DBLP:journals/misq/GefenKS03} & Model of trust formation in e-commerce settings. & None & None & \cite{DBLP:journals/ejis/SollnerHL16, DBLP:journals/jcis/HoehleHG12, Tam2020} & \cite{Troshani03092021}\\
    \hline
  \end{tabular}
\end{table*}

\begin{table}
  \caption{Included articles of our literature review; CR: critical review, Added: from search results of other discipline.}
  \label{tab:articles-literature-review}
  \vspace{-0.25\baselineskip}
  \setlength{\tabcolsep}{5pt}
  \centering
  \begin{tabular}{|p{0.11\linewidth}|p{0.25\linewidth}|p{0.14\linewidth}|p{0.09\linewidth}|p{0.085\linewidth}|}
    \hline \rowcolor{gray!40}
    \textbf{Area} & \textbf{Research\newline Articles} & \textbf{Literature\newline Reviews} & \textbf{Special\newline Issues} & \textbf{Added} \\
    \hline

    \textsc{HCI} & \cite{DBLP:journals/computer/SousaLCM24, BURGOON2000553, DBLP:journals/behaviourIT/Gulati0L19, DBLP:journals/ijhci/Shneiderman20, DBLP:conf/chi/AdarTT13, DBLP:journals/ijmms/RiegelsbergerSM05, DBLP:conf/um/KrausWM20, pinto2022trust, DBLP:journals/ijmms/CorritoreKW03a, DBLP:journals/chb/HuoZYSH22, Cai02012025, DBLP:journals/hhci/JoinsonRBS10, DBLP:conf/compsac/NothdurftBLM12} & \cite{DBLP:journals/ijhci/Jeon24, GULATI2024100495, DBLP:journals/ijhci/BachKHBS24, DBLP:journals/csur/MyersHC96} & None & \cite{DBLP:conf/chi/WidderDHHD21}\\
    \hline \rowcolor{gray!10}
    
    \textsc{HCI-AI} & \cite{DBLP:journals/ijhci/Shneiderman20, DBLP:journals/computer/SousaLCM24, DBLP:journals/chb/HuoZYSH22, DBLP:journals/ijhci/XuDGG23, DBLP:journals/behaviourIT/Gulati0L19, Cai02012025, DBLP:journals/ijhci/HuoZQYYYS24, DBLP:journals/pacmhci/BanovicYRL23, DBLP:journals/ijhci/SilvaSHGG23, DBLP:journals/ijhci/GaribayWAABCFFG23, DBLP:journals/chb/SuenH23, pinto2022trust, DBLP:journals/ijhci/YokoiEFN21, DBLP:conf/chi/AmershiWVFNCSIB19, shneiderman2020human} & \cite{Cai02012025, DBLP:journals/ijhci/BachKHBS24, 10.1145/3696449} & \cite{DBLP:journals/aisthci/RobertBMS20} & \cite{10.1145/3419764}\\
    \hline

    \textsc{IS} & \cite{DBLP:journals/jsis/LiHV08, DBLP:journals/ejis/SollnerHL16, DBLP:journals/tem/ThatcherMBAR11, DBLP:conf/hicss/ChopraW03, DBLP:journals/jcis/HoehleHG12, DBLP:journals/ejis/GrimsleyM07, Zhang2018, PARK2014153, DBLP:journals/isj/AllenCFK00, DBLP:journals/db/LiHV06, DBLP:journals/iam/MaoLD08, Tam2020, Tansley2007} & None & \cite{acc470ea-2310-31ca-a502-a1b6f480aafd, DBLP:journals/jmis/BenbasatGP08} & None \\
    \hline \rowcolor{gray!10}

    \textsc{IS-AI} & \cite{Saffarizadeh02072024, DBLP:journals/electronicmarkets/ThiebesLS21, DBLP:journals/jds/SchmidtBT20, Dennis03042023, https://doi.org/10.1111/jbl.12301, Das01112024, Troshani03092021, Korzyński12092024, DBLP:journals/giq/JanssenBEBJ20, DBLP:conf/wirtschaftsinformatik/TomitzaSSW23, info:doi/10.2196/15154} & \cite{DBLP:journals/csur/KaurURD23, DBLP:conf/hicss/LockeyGHS21, doi:10.5465/annals.2018.0057, DBLP:journals/electronicmarkets/YangW22} & \cite{DBLP:journals/electronicmarkets/LukyanenkoMS22} & \cite{DBLP:journals/isci/VirvouTT24} \\
    \hline
    
    \textsc{SE} & \cite{DBLP:journals/ese/BratthallJ02, DBLP:conf/esem/JalaliGS10, DBLP:conf/sigsoft/HassanLRGC00TOL24, DBLP:conf/icse/JohnsonBFFZ23, DBLP:journals/tse/LimamB10, DBLP:journals/software/BiancoLMT11, DBLP:journals/tse/CalinescuWGIHK18, DBLP:journals/jss/KeivanlooR14, DBLP:journals/ese/WangR16, DBLP:conf/icse/TrainerAR11, DBLP:conf/esem/BertramH0WS20, DBLP:journals/sigsoft/BeckerHPBKPDLRWGMSHMW06} & \cite{DBLP:journals/ese/HouJ23, DBLP:conf/sac/BuhnovaHIB23} & None & None \\
    \hline \rowcolor{gray!10}
    
    \textsc{SE-AI} & \cite{DBLP:journals/spe/AkbarKMRD24, DBLP:conf/icse/JohnsonBFFZ23} & \cite{DBLP:journals/csur/LiuLZJWS25, DBLP:journals/csur/ZhangDZ24, DBLP:journals/csur/KaurURD23, DBLP:journals/csur/LiQLDLPYZ23, DBLP:conf/hicss/LockeyGHS21} & \cite{DBLP:journals/electronicmarkets/LukyanenkoMS22} & None \\
    \hline
    
    \textsc{SE-CR} & \cite{DBLP:conf/icse/NollerS0R22, DBLP:journals/tse/AlamiPCW22, DBLP:journals/tse/WinterBCHHNW23, DBLP:journals/pacmse/KouCW0024, DBLP:conf/chi/BoggustHSS22} & None & None & None \\
    \hline\rowcolor{gray!10}
    
    \textsc{PH+PS} & \cite{DBLP:conf/fat/JacoviMMG21, DBLP:journals/tamd/HeGCMMM22, Yan2024, 10086944, li2024developing, Ryan2020AI, winfield2018ethical, DBLP:journals/ethicsit/Nickel22, pink2024trust, DBLP:conf/fat/0004CF024, chen2021trust, duran2025trust, mattioli2024overview, doi:10.1177/00218863241285033}  & None & None & None \\
    \hline
  \end{tabular}
\end{table}

\subsubsection{\textbf{Information Systems and Management (IS)}}
\label{sec:is}


In IS research, trust has been studied since the 1990s~\cite{sollner2013we}.
The articles we included in our review of IS literature were published in the time span from 2000 to 2024.
Of the 24 research papers we analyzed, only four did not embed their studies in existing trust models.
It was common for IS papers to customize existing trust models (13 out of 25 articles).

Several articles discussed that \enq{trust is a dynamic concept that develops over time}~\cite{DBLP:journals/jsis/LiHV08}.
The factors and processes that lead to \emph{trust formation} change gradually~\cite{Mayer1995}, and can include aspects such as people's perceptions about a company or brand~\cite{DBLP:journals/ejis/SollnerHL16}.
This means that trust in a company can transfer to trust in a particular tool, a concept called \emph{trust transference}~\cite{DBLP:journals/jmis/Stewart06, DBLP:journals/dss/WangSS13}.
\citeauthor{Saffarizadeh02072024} applied this concept to AI agents and noted that \enq{trust in an AI creator can be transferred to its AI agent when there is a perceived meaningful association between them}~\cite{Saffarizadeh02072024}.

IS articles clearly distinguish \emph{initial trust}, which is important to \enq{create a trustworthy first impression of a new system and encourage adoption}~\cite{DBLP:journals/jsis/LiHV08}, from the before-mentioned process of trust formation and later stages such as \emph{knowledge-based trust}, \enq{where the individual knows the other party well enough to predict the party's behavior in a situation}~\cite{DBLP:journals/electronicmarkets/ThiebesLS21}.
From a tool perspective, the later phases of trust are sometimes referred to as \enq{postadoption}, where aspects such as \emph{predictability} and \emph{helpfulness} of a tool~\cite{1985-30794-001} play an important role~\cite{DBLP:journals/tem/ThatcherMBAR11}.
\citeauthor{DBLP:journals/jsis/McKnightCK02}'s model is frequently referred to as a common IS model for initial trust, whereas \citeauthor{DBLP:journals/misq/GefenKS03}'s model focuses on trust formation.

In \citeauthor{DBLP:journals/jsis/McKnightCK02}'s model, based on \citeauthor{Mayer1995}, trust in vendors of e-commerce websites is a multidimensional construct centered around \emph{trusting beliefs}, that is, \enq{perceptions of the competence, benevolence, and integrity of the vendor} and \emph{trusting intentions}, that is, the \enq{willingness to depend} or, in other words, the \enq{decision to make oneself vulnerable to the vendor}~\cite{DBLP:journals/jsis/McKnightCK02}.
In addition to a trust model, \citeauthor{DBLP:journals/jsis/McKnightCK02} also introduced instruments to measure the concepts in their model (see \autoref{fig:mayer-scales} and \autoref{sec:se-adoption}). 

\citeauthor{DBLP:journals/misq/GefenKS03} models consumers' online purchase intentions as a combination of perceived usefulness, ease-of-use, and trust in the vendor~\cite{DBLP:journals/misq/GefenKS03}.
They show that online trust is built through \enq{(1) a belief that the vendor has nothing to gain by cheating, (2) a belief that there are safety mechanisms built into the web site, and (3) by having a typical interface, (4) one that is, moreover, easy to use.}

An important observation by \citeauthor{Mayer1995} is repeated in IS articles: humans differ in their general tendency to be willing to trust others~\cite{DBLP:journals/jsis/LiHV08}, that is, their \emph{propensity}~\cite{Mayer1995, DBLP:conf/hicss/ChopraW03} or \emph{disposition} to trust~\cite{DBLP:journals/db/LiHV06}.
These individual differences are a potential confounding factor in studies with developers investigating trust in AI assistants.

Two other central constructs are \emph{trusting intention}, which is \enq{the trustor's willingness to depend on the trustee}~\cite{DBLP:journals/jsis/LiHV08} and a person's \emph{trusting beliefs}~\cite{DBLP:journals/electronicmarkets/ThiebesLS21}, that is \enq{favorable object-specific beliefs}~\cite{DBLP:journals/tem/ThatcherMBAR11} that capture \enq{the trustor's perception of whether the trustee has the desired attributes to be trusted}~\cite{DBLP:journals/jsis/LiHV08} (see also \citeauthor{DBLP:journals/jsis/McKnightCK02}'s model).
Moreover, IS literature discusses how the trusting beliefs proposed by \citeauthor{Mayer1995} (benevolence, ability, and integrity) can foster technology acceptance~\cite{DBLP:journals/jais/BenbasatW05}.
\citeauthor{Troshani03092021} categorize those trusting beliefs into \emph{cognitive trust} and \emph{affective trust}~\cite{Troshani03092021} (see also \citeauthor{Komiak2006}~\cite{Komiak2006}).
Cognitive trust captures \enq{a customer's confidence or willingness to rely on a service provider's competence and reliability} while \emph{affective trust} is the \enq{confidence that one places in another party on the basis of feelings generated by the level of care and concern as demonstrated by the other party.}
In summary, competence (i.e., ability in \citeauthor{Mayer1995}'s model)  and predictability influence cognitive trust while benevolence and integrity influence affective trust~\cite{Troshani03092021}.

The IS literature discusses the validity of applying models of human trust to technology~\cite{DBLP:journals/jsis/LiHV08, DBLP:conf/hicss/ChopraW03, Saffarizadeh02072024}, which is referred to as \emph{plausibility of trust} in philosophy (see \autoref{sec:trust-and-ai}).
An aspect to consider is that humans have long been known to anthropomorphize technology~\cite{DBLP:journals/ijmms/NassMFRD95}.
To study trust relationships between humans and human-like tools, researchers have adapted the concept of \emph{interpersonal trust} to technology, for example, to study trust in automation or autonomous systems (see the description of \citeauthor{Lee2004Trust}'s trust model in \autoref{sec:hci}).
However, researchers have also argued that such models do not necessarily apply to GenAI tools due to their nondeterminism~\cite{info:doi/10.2196/15154}.

IS articles further discuss that increasing automation and complexity leads to systems being perceived as opaque and sophisticated, hence trust becomes more and more important~\cite{Lee2004Trust}.
This situation leads to uncertainty and a \emph{perceived risk} of using an automation tool, which could hinder adoption~\cite{DBLP:journals/jmis/BrownPR04, DBLP:journals/tem/ThatcherMBAR11}.
However, \citeauthor{DBLP:conf/hicss/ChopraW03} also write that \enq{trust can only arise when there exists a state of dependence between the trustor and trustee, and when acting on this dependence entails risk, i.e., the trustor possesses uncertainty about the outcomes and vulnerability to a potential loss if the outcomes are undesirable} \cite{DBLP:conf/hicss/ChopraW03}.
This is certainly true for software engineers who use GenAI tools to develop software.
The role of risk and vulnerability appears also in the psychology literature, in particular \citeauthor{Mayer1995} (see \autoref{fig:mayer-scales}), as well as in the philosophical debate summarized in \autoref{sec:philosophy}.

In the context of AI for decision-making, both \emph{mistakenly denied trust} and \emph{unfounded trust} can occur~\cite{DBLP:journals/jds/SchmidtBT20}. 
\citeauthor{DBLP:journals/jds/SchmidtBT20} summarize these situations as follows: \enq{If humans increasingly leverage AI to inform, derive, and justify decisions, it also becomes important to quantify when, how, why, and under which conditions they tend to overly trust or mistrust those systems}~\cite{DBLP:journals/jds/SchmidtBT20}.
This view corresponds to the concepts of \emph{undertrust} and \emph{overtrust} as described by \citeauthor{Lee2004Trust}~\cite{Lee2004Trust}.

Regarding \emph{trust quantification}, \citeauthor{DBLP:journals/jds/SchmidtBT20} note that \enq{trust is measured differently in different research fields}~\cite{DBLP:journals/jds/SchmidtBT20}.
However, they also mention that a common approach used across disciplines is to ask study subjects about their \emph{trusting beliefs} or \emph{trusting intentions} toward a certain entity, for which standardized survey instruments exist (cf. \cite{gefen2003managing}).
An exemplary operationalization of trust in decision-making based on observed behavior is calculating the probability that users follow a model's predictions by utilizing usage data~\cite{DBLP:conf/chi/Poursabzi-Sangdeh21, DBLP:journals/corr/abs-1901-08558}.
This resembles our motivating example of software developers accepting code suggestions.
We discuss trust operationalization and measurement in more detail in \autoref{sec:se-adoption}.

\subsubsection{\textbf{Human-Computer Interaction (HCI)}}
\label{sec:hci}

Trust has been studied in HCI since the late 1980s~\cite{DBLP:journals/ijmms/Muir87}.
Our review of HCI literature includes articles published between 1996 and 2024. 
Among the 24 unique HCI research papers we included, eight articles embedded their trust definitions in \citeauthor{Mayer1995}'s model. Three articles explicitly cite the definition, one paraphrases it, and four reference components such as \emph{ability}, \emph{integrity}, or \emph{benevolence} without adopting the complete model.
Furthermore, nine papers refer to \citeauthor{Lee2004Trust}'s trust model, and another five papers incorporate elements from both models.
Among the twelve papers building upon existing trust models, six extended the models either by developing new models or by adopting the definitions and further exploring specific aspects of trust. The remaining six papers refer to the models without modification.
In contrast, eleven papers do not explicitly refer to existing trust models.

Several HCI articles conceptualize trust as a \emph{context-dependent phenomenon} shaped by system behavior, user interaction, and situational factors. This perspective is especially prevalent in domains such as AI agents, automated decision support systems, and human-robot interaction. For example, \citeauthor{DBLP:journals/computer/SousaLCM24}~\cite{DBLP:journals/computer/SousaLCM24} propose the \emph{human-centered trustworthy framework}, which adapts \citeauthor{DBLP:journals/behaviourIT/Gulati0L19}'s human–computer trust model~\cite{DBLP:journals/behaviourIT/Gulati0L19} and integrates trustworthiness components from \citeauthor{Mayer1995}~\cite{Mayer1995}, along with constructs from \citeauthor{DBLP:journals/tmis/McKnightCTC11}~\cite{DBLP:journals/tmis/McKnightCTC11} and \citeauthor{Lee2004Trust}~\cite{Lee2004Trust}. In this framework, trust is approached as a multilevel concept that includes \emph{user predispositions}, \emph{interface features} such as transparency and control, and \emph{institutional factors} such as data governance and accountability. This layered view underscores the relational and evolving nature of trust in human–AI interaction. 

\citeauthor{pinto2022trust}~\cite{pinto2022trust} developed and validated a human–robot interaction trust scale by adapting \citeauthor{DBLP:journals/behaviourIT/Gulati0L19}'s model~\cite{DBLP:journals/behaviourIT/Gulati0L19}. Trust is conceptualized following \citeauthor{Mayer1995}~\cite{Mayer1995}'s definition as the \emph{willingness to be vulnerable to another party}, and is further informed by \citeauthor{Lee2004Trust}'s view of \emph{trust development over time}~\cite{Lee2004Trust}. Their adapted model incorporates dimensions such as \emph{reciprocity}, \emph{competence}, \emph{benevolence}, \emph{predictability}, \emph{honesty}, \emph{trust predisposition}, and \emph{willingness to be vulnerable}, tailored to human–robot interaction.

\citeauthor{DBLP:conf/compsac/NothdurftBLM12}~\cite{DBLP:conf/compsac/NothdurftBLM12} conceptualize trust as a multidimensional and dynamic construct that evolves through interaction. They draw on \citeauthor{Mayer1995}~\cite{Mayer1995} to describe trustworthiness in terms of \emph{ability}, \emph{integrity}, and \emph{benevolence}, and incorporate \citeauthor{Lee2004Trust}'s view of trust in automation~\cite{Lee2004Trust} as a \emph{calibrated attitude} shaped by system performance and user experience. Their conceptualization also reflects \citeauthor{madsen2000measuring}'s adaptation~\cite{madsen2000measuring} of \citeauthor{mcallister1995affect}'s definition~\cite{mcallister1995affect}, which emphasizes user confidence and willingness to act on the recommendations of intelligent systems. Building on this foundation, \citeauthor{DBLP:conf/compsac/NothdurftBLM12}~\cite{DBLP:conf/compsac/NothdurftBLM12} propose an adaptive explanation architecture that adjusts system transparency and feedback to support appropriate levels of trust in automated systems. 

\citeauthor{DBLP:journals/ijmms/RiegelsbergerSM05}~\cite{DBLP:journals/ijmms/RiegelsbergerSM05} conceptualize trust as a \emph{psychological state} characterized by a \emph{positive expectation} that one's \emph{vulnerability} will not be exploited.
Like \citeauthor{DBLP:conf/compsac/NothdurftBLM12}~\cite{DBLP:conf/compsac/NothdurftBLM12}, they build on \citeauthor{Mayer1995}~\cite{Mayer1995} and \citeauthor{mcallister1995affect}~\cite{mcallister1995affect}, but extend the discussion by emphasizing the \emph{social grounding} of trust.
Their framework is embedded in a trustor–trustee model adapted from the \emph{trust game}~\cite{2001-16661-005}, where trust is shaped by \emph{contextual cues}, \emph{reputation}, and \emph{internalized norms}.
They argue that trust in mediated interactions is not only influenced by \emph{individual traits} but also by \emph{perceived trustworthiness} of systems shaped through interface design and social context.

In contrast, several papers reference foundational trust models without adapting or extending them. \citeauthor{DBLP:journals/pacmhci/BanovicYRL23}~\cite{DBLP:journals/pacmhci/BanovicYRL23} cite \citeauthor{Mayer1995}~\cite{Mayer1995}, \citeauthor{mcallister1995affect}~\cite{mcallister1995affect} and \citeauthor{madsen2000measuring}~\cite{madsen2000measuring} to examine how untrustworthy AI systems can still elicit user trust, but do not propose a new framework or modify the referenced models.
Similarly, \citeauthor{DBLP:journals/ijhci/YokoiEFN21}~\cite{DBLP:journals/ijhci/YokoiEFN21} apply \citeauthor{Lee2004Trust}~\cite{Lee2004Trust} definition to investigate trust in AI-based medical decision-making, without extending it conceptually. \citeauthor{DBLP:journals/chb/SuenH23}~\cite{DBLP:journals/chb/SuenH23} adopt \citeauthor{Lee2004Trust}'s trust model to frame user trust in AI-based video interviews. Although the model is applied to a specific domain, its theoretical structure remains unchanged.
\citeauthor{DBLP:conf/um/KrausWM20}~\cite{DBLP:conf/um/KrausWM20} primarily build on \citeauthor{madsen2000measuring}~\cite{madsen2000measuring} trust model, while only briefly mentioning \citeauthor{Mayer1995} and \citeauthor{Lee2004Trust}.
These models are not embedded in their conceptual framework and rather serve as peripheral citations.
Similarly, \citeauthor{Cai02012025}~\cite{Cai02012025} adopt \citeauthor{Mayer1995}'s and \citeauthor{Lee2004Trust}'s definitions of trust to conceptualize user trust in human–computer interaction, but use the models primarily to validate the Chinese version of the human–computer trust scale in a new cultural context.

\subsubsection{\textbf{Software Engineering (SE)}}
\label{sec:se}

Trust is a critical element in SE that influences the adoption of software tools, processes, and collaborative practices.
However, our review of 13 articles published between 2002 and 2024 reveals that SE research typically does not engage with established foundational trust models.
In fact, only one paper~\cite{DBLP:conf/esem/JalaliGS10} cited \citeauthor{Mayer1995}'s trust definition and based their research on a related trust model.
Furthermore, one paper~\cite{DBLP:journals/ese/WangR16} implicitly referred to established trust models by mentioning \emph{risk}, that is, the aspect of \emph{vulnerability} (see \autoref{sec:psychology}).
The other papers either did not explicitly define trust or used custom definitions, diverging from disciplines such as IS and HCI, where established trust models are more commonly applied. 

An example of SE's customized approach to trust is the PICSE framework proposed by \citeauthor{DBLP:conf/icse/JohnsonBFFZ23}~\cite{DBLP:conf/icse/JohnsonBFFZ23}.
It describes a structured approach to understanding trust in software tools, identifying five key factors: \emph{personal}, \emph{interaction}, \emph{control}, \emph{system}, and \emph{expectation}.
The authors refer to \citeauthor{DBLP:conf/chi/FoggT99}'s credibility framework~\cite{DBLP:conf/chi/FoggT99} from HCI and \citeauthor{rousseau1998not}'s trust literature review~\cite{rousseau1998not} from IS.
However, these trust models did not contribute to PICSE but are discussed after introducing the framework.

\citeauthor{DBLP:conf/esem/JalaliGS10} adapted \citeauthor{DBLP:conf/pst/Schultz06}'s situational trust model~\cite{DBLP:conf/pst/Schultz06}, which extends \citeauthor{Mayer1995}'s model, to analyze trust dynamics in global software engineering~\cite{DBLP:conf/esem/JalaliGS10}.
In line with HCI and IS research, their model underlines the dynamic nature of trust, distinguishing \emph{initial trust building}, where trust forms based on past interactions, from \emph{trust evolution}, where trust fluctuates during the project based on expectations and behaviors.

\citeauthor{DBLP:journals/ese/WangR16} applied evolutionary game theory to examine trust and cooperation in distributed teams.
Their study does not explicitly model trust development or draw on established trust theories~\cite{DBLP:journals/ese/WangR16}.
Similarly, \citeauthor{DBLP:journals/tse/CalinescuWGIHK18} propose dynamic assurance cases for engineering trustworthy self-adaptive software, emphasizing system reliability and correctness rather than interpersonal trust~\cite{DBLP:journals/tse/CalinescuWGIHK18}.
\citeauthor{DBLP:journals/spe/AkbarKMRD24} take a different approach, presenting a taxonomy of decision-making challenges in trustworthy AI~\cite{DBLP:journals/spe/AkbarKMRD24}. 
Although these studies address trust-related concerns, they do so in domain-specific ways, emphasizing technical, ethical, or structural factors rather than explicitly modeling trust development based on established theories.

\citeauthor{DBLP:journals/ese/BratthallJ02} address trust in single data source software engineering case studies~\cite{DBLP:journals/ese/BratthallJ02}.
\citeauthor{DBLP:conf/esem/BertramH0WS20} investigate \emph{trust perceptions} in code review through an eye-tracking study, demonstrating how the perceived origin of code (human versus machine-generated) influences review behaviors and trust judgments~\cite{DBLP:conf/esem/BertramH0WS20}.
However, both papers do not explicitly reference or build upon established trust models.

\subsubsection{\textbf{Literature Reviews and Special Issues}}
\label{sec:literature-reviews-special-issue}

Among the included papers were 11 literature reviews. 
In addition, we found three editorials for special issues on trust.
Describing these articles in detail is beyond the scope of this paper.
However, we briefly summarize them because they reflect the scientific discourse on trust in the corresponding communities.

\fix{As part of our review, we found five literature reviews of trust that were \textbf{not limited to just one of our target disciplines} SE, HCI, or IS.}
\citeauthor{DBLP:journals/csur/KaurURD23}'s review centers around requirements for trustworthy AI~\cite{DBLP:journals/csur/KaurURD23}.
They refer to trust definitions across disciplines, including psychology, sociology, and economy, mentioning \emph{integrity} and \emph{reliability} as an \enq{agreement} across these disciplines.
They define trustworthy AI as a \enq{framework to ensure that a system is worthy of being trusted based on the evidence concerning its stated requirements}~\cite{DBLP:journals/csur/KaurURD23}.
\citeauthor{DBLP:journals/electronicmarkets/YangW22} examined 142 studies published between 2015 and 2022, identifying components, influencing factors, and outcomes of users' trust in AI~\cite{DBLP:journals/electronicmarkets/YangW22}.
\citeauthor{DBLP:journals/csur/LiQLDLPYZ23}'s review outlines aspects of AI trustworthiness such as \emph{robustness}, \emph{explainability}, \emph{transparency}, and \emph{fairness}, organized along the lifecycle of AI systems~\cite{DBLP:journals/csur/LiQLDLPYZ23}. However, they do not discuss established trust definitions or models.
\citeauthor{GULATI2024100495} reviewed 47 studies on trust in technology published in IS and HCI venues and found 17 different theories that were integrated from disciplines such as psychology and economics~\cite{GULATI2024100495}.
They conclude that \enq{the intricacies of how trust is formed and maintained in online environments still necessitates further investigation} and that the development of standardized and empirically validated instruments for trust measurement is crucial.
Finally, \citeauthor{DBLP:conf/hicss/LockeyGHS21}'s review focuses on \emph{antecedents of trust} in AI~\cite{DBLP:conf/hicss/LockeyGHS21}. They identify five AI-specific trust challenges: \emph{transparency and explainability}, \emph{accuracy and reliability}, \emph{automation versus augmentation}, \emph{anthropomorphism and embodiment}, and \emph{mass data extraction}~\cite{DBLP:conf/hicss/LockeyGHS21}.
An interesting observation regarding AI assistants for software development is that \enq{over-anthropomorphism may lead to overestimation of the AI's capabilities}~\cite{DBLP:conf/hicss/LockeyGHS21}, increasing risk~\cite{DBLP:journals/chb/CulleyM13}, decreasing trust~\cite{DBLP:journals/tiis/BakerPUK18}, and potentially leading to manipulation~\cite{Salles02042020}.
This is certainly an underexplored perspective on AI assistants for software development, especially when focusing on human factors of software security.

Three literature reviews focused on trust research in \textbf{HCI}.
\citeauthor{DBLP:journals/ijhci/Jeon24}'s review discussed \enq{the effects of emotions on trust in the context of technology use}~\cite{DBLP:journals/ijhci/Jeon24}.
Their review indicates that positive emotions lead to higher trust.
\citeauthor{DBLP:journals/ijhci/BachKHBS24} reviewed 23 empirical studies that focus on user trust in AI-enabled systems~\cite{DBLP:journals/ijhci/BachKHBS24}.
They describe trust definitions, factors that influence user trust, and measurement methods, reporting that seven studies explicitly defined trust, six of which referring to \citeauthor{Mayer1995}~\cite{Mayer1995} or \citeauthor{Lee2004Trust}~\cite{Lee2004Trust}.
Eight studies conceptualized trust, while nine studies neither defined nor conceptualized trust.
They found surveys, interviews, and focus groups to be the most commonly used methods to assess user trust.
\citeauthor{DBLP:journals/ijhci/BachKHBS24} conclude that it is important to select a definition of user trust that aligns with the specific study context.
\citeauthor{10.1145/3696449} reviewed 65 articles to assess the various definitions of appropriate trust in human-AI interaction, providing an overview of concepts and definitions and identifying similarities and differences between them~\cite{10.1145/3696449}.
They conclude that four common methods for building appropriate trust are improving \emph{system transparency}, considering user \emph{cognition and perception}, using \emph{guidelines or models} to achieve calibrated trust in AI, and considering the entire \emph{continuum of trust}, including \emph{over-, under-, mis-, and dis-trust}.

One literature review had a clear \textbf{IS} focus.
\citeauthor{doi:10.5465/annals.2018.0057} synthesized two decades of empirical research on the determinants of human trust in AI across disciplines~\cite{doi:10.5465/annals.2018.0057}.
The authors motivate how trust in AI differs from other technologies and identify an AI's capabilities as an important antecedent of trust development.
In addition, they propose a framework that identifies factors that influence users' \emph{cognitive} and \emph{emotional} trust.
Regarding the development of \emph{cognitive trust}, they highlight the importance of an AI's \emph{tangibility}, \emph{transparency}, \emph{reliability}, and \emph{immediacy}.
For \emph{emotional trust}, on the other hand, and AI's \emph{anthropomorphism} is a major factor.
\citeauthor{doi:10.5465/annals.2018.0057} also critically reflect on past studies by pointing to the diversity of trust measure used and the lack of longer-term studies in \enq{higher stakes environments}~\cite{doi:10.5465/annals.2018.0057}.

For IS, we further identified three special issues on trust.
One of them was published in 2010 in their flagship journal \emph{MIS Quarterly}, focusing on \enq{novel perspectives} on trust in IS.
The fact that this special issue was published 15 years ago is another indication of the maturity of trust research in that discipline.
The other IS-specific special issues focused on trust in online environments~\cite{DBLP:journals/jmis/BenbasatGP08} and trust in AI~\cite{DBLP:journals/electronicmarkets/LukyanenkoMS22}.

For \textbf{SE}, we included two literature reviews.
\citeauthor{DBLP:journals/csur/LiuLZJWS25} conducted a tertiary study reviewing 141 secondary studies on trustworthy AI and software, describing trustworthiness as a \enq{highly abstract concept comprising related quality attributes}~\cite{DBLP:journals/csur/LiuLZJWS25}. The review focuses on these quality attributes and compares them between trustworthy AI and other trustworthy software.
\citeauthor{DBLP:journals/csur/LiuLZJWS25} do not focus on the conceptualization of trust and trustworthiness.
\citeauthor{DBLP:journals/ese/HouJ23} reviewed 112 articles on trust in software ecosystems, identifying different definitions of concepts such as \emph{software trust}, \emph{system trust}, or \emph{software service trust}. 
However, their focus was on compiling trust definitions, not unifying or aligning them~\cite{DBLP:journals/ese/HouJ23}.

\begin{results}
\fix{\textbf{RQ2} \textsc{(trust conceptualizations)}:\\
In \textbf{information systems} (IS), trust has been studied since the 1990s; in \textbf{human-computer interaction} (HCI) since the late 1980s.
Both IS and HCI research is usually embedded in existing trust models (see \autoref{tab:trust-models}).
The \citeauthor{Mayer1995} (\citeyear{Mayer1995}) model is common across IS and HCI, \citeauthor{Lee2004Trust} (\citeyear{Lee2004Trust}) is more common in HCI.
Customizing or extending existing trust models is common in both disciplines.
Important concepts mentioned in IS research are \emph{initial trust}, \emph{trust formation}, and \emph{trust transference} (e.g., trust in a company transferring to a tool).
Trust is sometimes studied in the context of \emph{tool adoption} and \emph{technology acceptance}, but as one of several factors.
IS research commonly adopts or develops \emph{validated instruments} to measure trust-related concepts.
HCI research frames trust as a \emph{context-dependent} phenomenon.
\textbf{Software engineering} (SE) research typically does not engage with established trust models; trust as a concept is either not or only informally defined.}
\end{results}

\subsection{Critical Review of Trust Conceptualizations in Software Engineering Research}
\label{sec:criticalreviewinse}

Of the 58 articles that we analyzed as part of our critical review, only one article referred to an established trust model, and only three articles provided trust definitions.
This underlines the problematic situation suggested by our literature across disciplines (see \autoref{sec:se}).
There is a significant gap in the maturity of trust research in SE compared to IS and HCI.
Trust as a concept is often discussed in a context-specific manner without embedding concepts and definitions in established trust theories or models.
This lack of embedding in established trust theories from other disciplines hinders interdisciplinary comparisons and research toward a deeper understanding of what constitutes trust in SE-specific settings.

\citeauthor{DBLP:conf/icse/NollerS0R22} are the only authors who explicitly reference an established trust model, incorporating \citeauthor{Lee2004Trust}'s model of human-automation trust~\cite{DBLP:conf/icse/NollerS0R22}.
\citeauthor{DBLP:journals/tse/AlamiPCW22} define trust as \enq{the unyielding belief that the person is truthful and reliable} and in the context of their research as \enq{the willingness of the community to rely on the contributor, a prerequisite for considering her code change}~\cite{DBLP:journals/tse/AlamiPCW22}. This conceptualization frames trust as a social and reputation-based construct, essential in collaborative environments such as open-source software development, where trust is built through ongoing contributions.
Although this framing is reasonable, an embedding in existing trust models would allow researchers to identify the particularities of trust in open-source software compared to other settings.

\citeauthor{DBLP:journals/tse/WinterBCHHNW23}~\cite{DBLP:journals/tse/WinterBCHHNW23} implicitly define trust by referencing dispositional trust as proposed by \citeauthor{DBLP:journals/hf/HoffB15}~\cite{DBLP:journals/hf/HoffB15}, which describes an individual's inherent tendency to trust automation, independent of context or system performance. This positions trust as a psychological trait that varies between individuals.
However, it is important to note that trust was only a peripheral topic, not a central focus of their study.
\citeauthor{DBLP:journals/pacmse/KouCW0024} discuss trust in the context of large language models (LLMs) for code generation, suggesting that users assess trustworthiness based on the model's explainability and how well its predictions align with their expectations~\cite{DBLP:journals/pacmse/KouCW0024}.
However, they neither explicitly define trust nor consider the broader explainability discussion (see \autoref{sec:foundationoftrust}). Instead, they cite previous work by \citeauthor{DBLP:conf/chi/BoggustHSS22}, who explored how users determine whether to trust a machine learning model based on its ability to convey behavior in a way that aligns with their expectations~\cite{DBLP:conf/chi/BoggustHSS22}. Neither study provides a formal definition of trust; rather, both frame it in terms of user perceptions and interpretability.
Given the amount of related work on trustworthiness and explainability of AI-based systems in philosophy (see \autoref{sec:trust-and-ai}) and IS (see \autoref{sec:is}), it is surprising that this article is again not broadly embedded in established trust conceptualizations.

\begin{results}
\fix{\textbf{RQ2} \textsc{(trust in SE research)}:\\
Our critical review confirmed that there is a \textbf{significant maturity gap} in trust-related research in SE compared to IS and HCI.
Of the 58 analyzed articles, only one referred to an established trust model, and only three provided trust definitions.
This lack of embedding in established trust models hinders interdisciplinary comparisons and research toward a deeper understanding of what constitutes trust in SE-specific settings.}
\end{results}

\section{Adopting Trust Models in SE (RQ3)}
\label{sec:se-adoption}


To answer \textbf{RQ3}, we first discuss problems with the way SE research currently handles trust research and then describe how the common trust model of \citeauthor{Mayer1995}~\cite{Mayer1995} could be adopted to study trust in AI assistants for software development.

\subsection{Current State of Trust Research in SE}

Our research was motivated by the fact that several SE research articles suggested approaches to increase `trust' with the aim of increasing tool adoption (see \autoref{sec:introduction}).
Although previous research in other fields such as IS suggests that trust is indeed related to tool adoption and usage (see \autoref{sec:is}), the corresponding research articles thoroughly define and conceptualize trust and related concepts such as \emph{trust transference}, \emph{trusting beliefs}, or a user's \emph{propensity to trust}.
These conceptualizations originate in mature trust models, in particular the one proposed by \citeauthor{Mayer1995} in \citeyear{Mayer1995}~\cite{Mayer1995}.
However, in terms of trust and AI, \citeauthor{DBLP:journals/electronicmarkets/ThiebesLS21} note that \enq{trust in general is a complex phenomenon that has sparked many scholarly debates in recent decades}, therefore \enq{it is not surprising that the conceptualization of trust in AI and what makes AI trustworthy---as of today---remains inconclusive and highly discussed in research and practice}~\cite{DBLP:journals/electronicmarkets/ThiebesLS21}.
It is unfortunate that the SE research community is not a major contributor to that interdisciplinary discourse because, as our review of SE literature has shown, SE research (and practice, see \cite{pink2024trust}) on trust mainly uses ad hoc trust definitions without embedding studies in established trust models.

\renewcommand{\arraystretch}{1.2}
\begin{table}
    \caption{Summary of the tenets of trust as discussed in philosophy and psychology compared to software engineering, where trust has been operationalized as artifact acceptance.}
    \label{tab:tenets-trust}
    \centering
    \small
    \begin{tabular}{|p{0.47\linewidth}|p{0.45\linewidth}|} \hline \rowcolor{gray!50}
        \textbf{Tenets of trust in philosophy (\textsc{ph}) and psychology (\textsc{ps})} & \textbf{Operationalizing trust as artifact acceptance...}\\ \hline 
        Trust is an attitude, not a behavior (\textsc{ph}, \textsc{ps}). & ...treats trust as a behavior.\\ \hline \rowcolor{gray!10}
        Trust is context-dependent\newline (\textsc{ph}, \textsc{ps}). &...just focuses on one context.\\ \hline 
        Trust should be well-calibrated or well-grounded (\textsc{ph}, \textsc{ps}). &...does not measure calibration or grounding.\\ \hline \rowcolor{gray!10}
        Trust should have a high resolution (\textsc{ps}). &...does not measure resolution.\\ \hline 
        Trust should have a high specificity (\textsc{ps}). &...could at most measure low specificity.\\ \hline \rowcolor{gray!10}
        Trust should be justified (\textsc{ph}). &... ignores justification.\\ \hline 
        Trust should be plausible (\textsc{ph}). &... ignores plausibility.\\ \hline \rowcolor{gray!10}
    \end{tabular}
\end{table}

The detachment of SE research from established trust models is problematic, as evidenced by the equation of trust with the acceptance of generated content, which is shortsighted for several reasons. 
Drawing on the foundations of trust from psychology and philosophy that we introduced in \autoref{sec:foundationoftrust}, \autoref{tab:tenets-trust} shows how such an operationalization disregards the tenets of trust discussed in these disciplines. 
Most importantly, both disciplines see trust as an \emph{attitude} and not a \emph{behavior}.
Therefore, measuring acceptance disregards factors that influence a person's behavior, such as their situation (e.g., time pressure), the type of AI system they are interacting with, their specific task, the potential consequences of decisions, and many more.
Still, even if acceptance measured trust, it would not measure \emph{calibrated} or \emph{warranted trust}.
Hence, the measured trust could be completely detached from the actual capabilities of an AI assistant.
However, given the fact that GenAI-based software development assistants will continue to hallucinate~\cite{DBLP:journals/corr/abs-2409-05746}, the potentially resulting maintainability challenges~\cite{DBLP:journals/tosem/LiuLWTLLL24}, \fix{security risks~\cite{DBLP:conf/ccs/PerryS0B23, DBLP:conf/sp/PearceA0DK22, DBLP:conf/uss/SandovalPNKGD23, DBLP:conf/chi/SerafiniYN25, DBLP:conf/sac/KudriavtsevaHG25, DBLP:conf/wcre/MajdinasabBRDTK24, DBLP:conf/ccs/KlemmerHPLBPMRV24, DBLP:conf/sp/OhLPKK24}}, and negative impact on learning~\cite{Yan2024, DBLP:journals/cacm/Shein24b}, it would not only be morally but also pragmatically better to find out whether developers' trust in such tools is actually calibrated or warranted.
After all, maintainability issues or security vulnerabilities introduced by relying on generated code could cause major financial and reputational damage to individuals and companies. 

It was interesting for us to see that IS and HCI primarily focus on the psychological side of trust.
We posit that discussion in SE can also benefit from incorporating philosophical insights on trust.
For example, the discussion whether trust in AI is even conceptually possible is relevant in the sense that it critically questions whether engineering for trust is not just a marketing phenomenon or, in the worst case, even ethicswashing (see \autoref{sec:trust-and-ai}).
A similar phenomenon can be observed in the debates on XAI, where the expectation is that XAI invariably leads to greater trust, something that could not be proven in studies~\cite{Kaestner2021Relation}. In this sense, the concept of \emph{justification} is also important for SE, as it can help ensure that trust in AI assistants for software development is formed appropriately, rather than, for example, being the result of misleading marketing rhetoric.
It is surprising that the trust conceptualizations in SE are still so underdeveloped, as tools from SE can be extremely important for creating trustworthy systems. For example, \citeauthor{DBLP:conf/ecai/AhujaBBCGLMSSS20} argue that SE processes and tools, in particular a detailed requirements analysis, monitoring instruments, and automated testing, are essential to create systems that can really be trusted~\cite{DBLP:conf/ecai/AhujaBBCGLMSSS20}.

\subsection{\fix{General Role of Trust in SE}}

\fix{In software development, negative outcomes, such as bugs or security vulnerabilities introduced by using AI assistants, materialize with a certain delay, and are often borne by someone else.
Human reviewers must invest more time reviewing large amounts of generated code, security teams must catch vulnerabilities in production, bugs slow users down, and the organization may face a reputation loss due to data leakage or outages.
This creates an incentive landscape in which an individual developer can rationally feel a low personal risk even when the system-level risk is unchanged or even increased.
In trust terms, this can lead to an unjustifiably low perceived risk and potentially even to an unjustifiably high trust propensity.}

\fix{Since developers are rewarded for throughput (e.g., resolved tickets), and costs of defects accumulate as technical debt, teams might drift into thinking (and trusting) that ``it usually works.''---while long-term maintenance costs are accumulating.
This makes trust in software contexts unusually sensitive to incentives, governance, and feedback frequency.
Such factors are different in settings where the consequences are immediate and personal.
In SE, however, a developer might have already left the company before the problems, which their trust in AI assistants has caused, materialize.}

\fix{Moreover, the object that is trusted can shift in an SE context.
Software development is embedded in institutionalized safety nets that involve code reviews, software tests, and compliance checks, often automated or supported by continuous integration pipelines.
These processes can be risk-reducing, so developers may come to trust the pipeline rather than the AI assistant (\enq{our tests will catch potential issues}).
This mindset can lead to \enq{trust laundering}, where weak trust in an AI output is converted into or substituted by greater trust in the infrastructure.
The effect of this is that trust becomes procedural and distributed.
It is not \enq{I trust the AI assistant} but \enq{I trust our socio-technical infrastructure to detect and absorb errors.}}

\fix{Another philosophical and legal issue is that AI assistants lead to ambiguous authorship and accountability. When AI suggestions are integrated, it can become even more unclear who is responsible for an issue: Is it the developer who used the AI assistant, the} \fix{reviewer who approved the change, or the team whose tests did not catch the bug?
If accountability becomes more diffuse, developers may feel less personally exposed, which again raises trust even though organizational risk remains the same.
In contrast, in teams with strong code ownership, norms, and guidelines, the same ambiguity can reduce trust because accepting AI output threatens clear boundaries of responsibility.
One could argue that trust really depends on the institutional practices in place.}

\subsection{An Example of Adopting Trust Models in SE}

The dimensions of perceived trustworthiness proposed by \citeauthor{Mayer1995}~\cite{Mayer1995}, that is, \emph{ability} (A), \emph{benevolence} (B), and \emph{integrity} (I) can be easily transferred to AI assistants for software development.
As mentioned before, \emph{ability} requires AI assistants to generate correct and reliable code that is appropriate for a given context.
\emph{Benevolence} reflects interaction patterns that AI vendors implement to ensure that AI assistants are perceived as supportive, and \emph{integrity} models the data curation and guardrails that AI vendors and regulatory bodies implement to avoid biases or problematic code that, for example, contains security vulnerabilities.
In the following, we discuss existing instruments that SE researchers can use, with a particular focus on our own adaption of \citeauthor{Mayer1995}'s trust model as presented in \autoref{fig:mayer-scales}. 
Please note that this is one example for adapting an established trust model to an SE context.
If concepts such as \emph{calibrated trust} are important for a particular SE research project, the model by \citeauthor{Lee2004Trust} might be more suitable~\cite{Lee2004Trust}.
It is important that SE researchers know enough about trust conceptualizations to understand the implications of adapting particular trust models themselves.
We hope that our article serves as a starting point for this.

As already mentioned, operationalizing trust is difficult; one needs to be specific about the actual concepts that one intends to capture (e.g., the \emph{perceived trustworthiness} of a tool or a person's general \emph{propensity to trust}).
Fortunately, validated instruments exist to assess \citeauthor{Mayer1995}'s ABI dimensions of perceived trustworthiness.
SE researchers can adapt these instruments to study trust in AI assistants~\cite{MayerDavies1999, DBLP:journals/isr/McKnightCK02}.
There further exist various instruments to assess a person's \emph{propensity to trust}~\cite{YamagishiYamagishi1994, DBLP:journals/isr/McKnightCK02, Frazier01102013, Scholz17012025}.
Since perceived \emph{risk} and \emph{vulnerability} are central to \citeauthor{Mayer1995}'s model, we also want to refer to instruments designed to measure related concepts such as a person's \emph{willingness to be vulnerable} or their \emph{trusting intensions}~\cite{DavidSchoorman02012016, DBLP:journals/isr/McKnightCK02, gillespie201520}.

As mentioned above, IS literature has discussed operationalizing trust based on observed behavior by calculating the probability that users follow a model's prediction based on usage data~\cite{DBLP:conf/chi/Poursabzi-Sangdeh21, DBLP:journals/corr/abs-1901-08558}.
Since this is closely related to our motivating example of software developers accepting code suggestions, we want to discuss how such metrics can be included in our representation of \citeauthor{Mayer1995}'s trust model (\autoref{fig:mayer-scales}).
\citeauthor{DBLP:journals/jds/SchmidtBT20} argue that behavioral metrics might work well as proxies of trust when comparing different treatments (e.g., different ways of presenting AI suggestions) or comparing different groups based on the same treatment~\cite{DBLP:journals/jds/SchmidtBT20}.
However, factors such as participant knowledge and confidence, and (perceived) task difficulty are known to affect advice acceptance~\cite{YANIV20041}.
Over time, the usage of tools moderates users' bias toward self-reliance in decision-making and could further influence the observed behavior~\cite{DBLP:journals/ijmms/DzindoletPPPB03}.

In the context of \citeauthor{Mayer1995}'s model, the acceptance rate for AI suggestions can be interpreted as an indicator of the tool's ``observed outcome.'' 
Studying such data in detail is crucial, as acceptance of AI suggestions can be a factor by which people form their trust; that is, in \citeauthor{Mayer1995}'s model, acceptance rates can contribute to the \emph{perceived ability} of the system.
Support for this view comes, for example, from \citeauthor{DBLP:conf/fat/0004CF024}'s study.
They have developed and tested design concepts, including a dashboard showing a user's personal GitHub Copilot acceptance rate, finding that it helped developers align their expectations with an AI assistant's ability~\cite{DBLP:conf/fat/0004CF024}.

A general limitation of the questionnaire-based instruments referenced above is that answering survey questions can never fully capture trust as conceptualized by \citeauthor{Mayer1995}.
They cannot capture critical situations in which a person actually made itself \enq{vulnerable to the actions of another party}~\cite{Mayer1995} (here: an AI assistant).
As mentioned above, proxies of \emph{behavioral trust} can complement questionnaire-based instruments.
However, they also cannot capture actual behavioral trust.
A general example of a situation in which behavioral trust occurs would be that parents leave their children with a certain babysitter as an indicator of trusting the babysitter~\cite{DBLP:journals/jds/SchmidtBT20}.
Another limitation is that human behavior and decision-making are complex and can be affected by factors such as monetary constraints or a lack of alternatives.
Therefore, researchers must be careful when using behavior as a proxy for trust, considering that there might be different drivers for the observed behavior.

\subsection{Recommendations}

Our general recommendation is that SE researchers should keep in mind that focusing on individual factors such as \emph{propensity to trust} or \emph{behavioral metrics} is generally not sufficient to capture a holistic notion of trust.
An approach to mitigating threats to validity is to combine different factors and instruments, ideally based on an established trust model.
We want to conclude this section with a list of actionable recommendations for researchers who want to conduct studies on trust in an SE context.
These recommendations are based on our thorough description of the foundations of trust (\autoref{sec:foundationoftrust}) and our literature review across disciplines (\autoref{sec:trustacrossdisciplines}):

\begin{enumerate}[leftmargin=*]
    \item Before designing concrete studies on trust in AI assistants, carefully read the literature on \textbf{trust beyond SE}. We hope that our article serves as a useful starting point for this.
    \item When reporting study design and results in a paper, embed them in \textbf{established trust models} (see \autoref{tab:trust-models}). 
    \item Use \textbf{established terminology} when discussing trust, for example, distinguishing \emph{overtrust}/\emph{undertrust} from \emph{calibrated trust} or distinguishing \emph{initial trust} from \emph{trust formation} (see \autoref{sec:foundationoftrust} and \autoref{sec:is}).
    \item Do not equate artifact acceptance with trust. When designing empirical studies to assess developers' trust in AI assistants, clearly outline which \textbf{aspects of trust} are being measured using which instruments (e.g., \emph{cognitive trust} versus \emph{affective trust}). Use \textbf{validated instruments} whenever possible (see trust operationalizations discussed in this article) and openly discuss their limitations.
    \item Consider participants' \emph{trust disposition} and \emph{trusting beliefs} as \textbf{confounding factors}; consider the \emph{transferable} nature of trust. For example, trust in a particular AI vendor can propagate to a specific AI assistant that this vendor offers.
    \item Clearly describe the \textbf{study context} and reflect on its generalizability. Given how central \emph{risk} and being \emph{vulnerable} are for trusting behavior, a study in a low risk scenario might not generalize to a high risk situation, for example, in the context of safety-critical software.
\end{enumerate}

\begin{results}
\fix{\textbf{RQ3} \textsc{(Recommendations for SE)}:\\
SE researchers should treat trust as a \emph{multidimensional} and \emph{context-dependent} phenomenon.
Focusing on individual factors is generally not sufficient to holistically capture trust.
Therefore, SE researchers should build on established \emph{trust models} and \emph{measurement instruments}.
Section~IV-D provides further recommendations.}
\end{results}

\section{Threats to Validity}
\label{sec:validity}
As with any empirical study, our methodology is subject to certain limitations that may affect the completeness, accuracy, and generalizability of our findings. In the following, we identify and discuss potential threats to the validity of our methodology, including biases in paper selection, interpretation of trust conceptualizations, and the applicability of findings across disciplines. We classify the validity threats in three categories as proposed by~\citeauthor{wohlin2012experimentation}~\cite{wohlin2012experimentation}.

\emph{Internal Validity} concerns whether our study design and execution introduced biases that may affect the correctness of our findings. A potential threat to internal validity is that we may have missed indirect or implicit definitions of trust as well as references to trust models during our literature review. Additionally, by focusing on the highest-ranked articles from our queries, we may have unintentionally excluded relevant papers ranked lower. We acknowledge the selective inclusion based on an opaque ranking algorithm provided by Google Scholar as a limitation.
However, Google Scholar is used by academics across disciplines and allowed us to use the same sampling approach for all disciplines.
\fix{We complemented the review of SE literature based on Google Scholar search results by conducting a critical review of recently published SE papers, confirming our earlier results.}

Future work should extend the literature review by using discipline-specific retrieval approaches.
\fix{One related discipline that studies AI assistants for software development~\cite{DBLP:conf/ccs/PerryS0B23, DBLP:conf/sp/PearceA0DK22, DBLP:conf/uss/SandovalPNKGD23}, sometimes referring to human trust~\cite{DBLP:conf/sp/OhLPKK24, DBLP:conf/ccs/KlemmerHPLBPMRV24}, is the security research community.
We did not explicitly exclude security-related research in our review, only mentions of trust in a more technical context, as clarified in \autoref{sec:introduction}.
However, at the same time, we did not explicitly include security research venues.
Extending the literature review to consider security research as well is a promising direction for future work.}

\emph{External Validity} concerns the generalizability of our findings. Our literature review focused on SE, HCI, and IS; our description of the foundations of trust is rooted in philosophy and psychology.
We did not cover trust conceptualizations in other disciplines, such as sociology.
However, given that our literature review primarily revealed trust models rooted in psychology, these foundations seem to be the most relevant ones for the considered disciplines.
As we only considered English-language publications, our findings may not reflect cultural variations in trust.

\emph{Construct Validity} concerns whether we accurately captured and analyzed the concept of trust as intended.
Given the complexity of trust and its multidimensional nature, one potential threat is our own limited understanding of the concept.
To mitigate this, we recruited coauthors from philosophy and HCI and collected feedback from another colleague with a background in psychology and IS.

\section{Conclusion}
\label{sec:conclusion}

Our review of SE, IS, and HCI literature has shown that SE articles, contrary to IS and HCI, often use the term `trust' informally and that providing an explicit definition of trust or embedding results in established trust models is rare.
However, without a common definition, true secondary research on trust is impossible.
In this paper, we have discussed the consequences of the lack of common trust models in SE research, and how the SE research community can learn from other disciplines.
In \autoref{sec:se-adoption}, we present an example of an adaption of an established trust model in an SE context and provide actionable recommendations for SE researchers planning to study trust in AI assistants.
Investigating the relationship between trust-related concepts for which established instruments exist and behavioral metrics such as artifact acceptance in the context of AI assistants for software development is a promising direction for future work.
We hope that our paper contributes to more mature and interdisciplinarily embedded SE trust research in the future.


\balance
\printbibliography

\end{document}